\documentclass[conference,compsoc]{IEEEtran}
\ifCLASSOPTIONcompsoc
  \usepackage[nocompress]{cite}
\else
  \usepackage{cite}
\fi
\ifCLASSINFOpdf
\else
\fi
\usepackage{tikz}
\usepackage{amsmath}

\usepackage{filecontents}
\usepackage{amsthm}
\usepackage{comment}
\usepackage{threeparttable}
\usepackage{multirow}
\usepackage{titlesec}
\usepackage{multicol}
\theoremstyle{definition}

\makeatletter
\def\thm@space@setup{\thm@preskip=1pt
\thm@postskip=1pt}
\makeatother

\usepackage{graphicx}
\usepackage{algorithm}
\usepackage{amsmath}
\usepackage{algpseudocode}

\usepackage{tikz}
\usepackage{amsfonts}
\usepackage{ifthen}

\usepackage{enumitem}

\usepackage{booktabs}

\usepackage{mdframed}
\mdfsetup{skipabove=0.5pt,skipbelow=0.5pt}
\newmdenv{allfour}
\newmdenv[leftline=false,rightline=false, linecolor=darkgray, linewidth = 0.2mm, startinnercode={\baselineskip=0cm}]{topbot}
\newmdenv[topline=false,rightline=false]{leftbot}
\newmdenv[topline=false,bottomline=false]{leftright}

\usepackage{subcaption}
\usepackage{gensymb}
\usepackage{pifont}
\usepackage{makecell}

\usepackage{booktabs}
\usepackage{moresize}
\usepackage{tabularx}

\usepackage{tabularray}
\definecolor{Silver}{rgb}{0.752,0.752,0.752}

\newboolean{commentsOn}
\setboolean{commentsOn}{true}

\newcommand{\mr}[1]{{\color{black}#1}}

\usepackage{xspace}

\usepackage[scaled=.75]{beramono}

\def\eg{{e.g.},~}

\usepackage{microtype}
\usepackage{amsmath,stackengine}
\stackMath
\usepackage{ifthen}

\newcounter{sqindex}

\newcommand{\shortsectionBf}[1]{\vspace{3pt}
\noindent {\bf #1}}

\usepackage{setspace} 

\newmdenv[backgroundcolor=gray!5,
    linecolor=gray,
    outerlinewidth=1pt,
    roundcorner=5,
    skipabove=2pt,
    skipbelow=0pt,
    innerleftmargin=3pt,
    innerrightmargin=3pt,
    innerbottommargin=3pt,
    innertopmargin=3pt,
    font=\normalsize
]{shadedboxed}

\PassOptionsToPackage{hyphens}{url}
\usepackage{hyperref}
\mathchardef\mhyphen="2D
\usepackage{wasysym}

\makeatletter
\def\thickhline{%
  \noalign{\ifnum0=`}\fi\hrule \@height \thickarrayrulewidth \futurelet
   \reserved@a\@xthickhline}
\def\@xthickhline{\ifx\reserved@a\thickhline
               \vskip\doublerulesep
               \vskip-\thickarrayrulewidth
             \fi
      \ifnum0=`{\fi}}
\makeatother

\newlength{\thickarrayrulewidth}
\usepackage{listings}
\usepackage{color, colortbl}
\usepackage{tikz}
\definecolor{light-gray}{gray}{0.95}

\colorlet{punct}{red!60!black}
\definecolor{background}{HTML}{EEEEEE}
\definecolor{delim}{RGB}{20,105,176}
\colorlet{numb}{magenta!60!black}

\lstdefinelanguage{json}{
    basicstyle=\footnotesize\ttfamily,
    numbers=left,
    numberstyle=\scriptsize,
    stepnumber=1,
    numbersep=2pt,
    showstringspaces=false,
    breaklines=true,
    frame=lines,
    xleftmargin=1em,
    backgroundcolor=\color{background},
    literate=
     *{0}{{{\color{numb}0}}}{1}
      {1}{{{\color{numb}1}}}{1}
      {2}{{{\color{numb}2}}}{1}
      {3}{{{\color{numb}3}}}{1}
      {4}{{{\color{numb}4}}}{1}
      {5}{{{\color{numb}5}}}{1}
      {6}{{{\color{numb}6}}}{1}
      {7}{{{\color{numb}7}}}{1}
      {8}{{{\color{numb}8}}}{1}
      {9}{{{\color{numb}9}}}{1}
      {:}{{{\color{punct}{:}}}}{1}
      {,}{{{\color{punct}{,}}}}{1}
      {\{}{{{\color{delim}{\{}}}}{1}
      {\}}{{{\color{delim}{\}}}}}{1}
      {[}{{{\color{delim}{[}}}}{1}
      {]}{{{\color{delim}{]}}}}{1},
}

\usepackage{tikz}
\usepackage{xcolor}

\usepackage{etoolbox} 
\AtBeginEnvironment{quote}{\vspace{0.75em}}
\AtEndEnvironment{quote}{\vspace{0.75em}}

\begin{document}
%
\title{{\fontsize{11}{12}\selectfont \textnormal{ Accepted to IEEE Symposium on Security and Privacy 2026}} \\[1ex]  International Students and Scams: At Risk Abroad
\vspace{-1.5em}
}

\newcommand\copyrighttext{
  \footnotesize \textcopyright 2025 IEEE. Personal use of this material is permitted. Permission from IEEE must be obtained for all other uses, in any current or future media, including reprinting/republishing this material for advertising or promotional purposes, creating new collective works, for resale or redistribution to servers or lists, or reuse of any copyrighted component of this work in other works.
}
\newcommand\copyrightnotice{
  \begin{tikzpicture}[remember picture,overlay]
    \node[anchor=south,yshift=10pt] at (current page.south)
    {\fbox{\parbox{\dimexpr\textwidth-\fboxsep-\fboxrule\relax}{\copyrighttext}}};
  \end{tikzpicture}
}

\author{
{\rm Katherine Zhang$^\dagger$, Arjun Arunasalam$^\ddagger$, Pubali Datta$^\dagger$, and Z.\ Berkay Celik$^\mathsection$} \\
$^\dagger$ University of Massachusetts Amherst, \{kzhang, pdatta\}@umass.edu \\
$^\ddagger$ Florida International University, aarunasa@fiu.edu \\
$^\mathsection$ Purdue University, zcelik@purdue.edu
}


%


\maketitle
\copyrightnotice

\vspace{-1.25em}
\begin{abstract}
International students (IntlS) in the US refer to foreign students who acquire student visas to study in the US, primarily in higher education. 
As IntlS arrive in the US, they face several challenges, such as adjusting to a new country and culture, securing housing remotely, and arranging finances for tuition and personal expenses.
These experiences, coupled with recent events such as visa revocations and the cessation of new visas, compound IntlS' risk of being targeted by and falling victim to online scams.
While prior work has investigated IntlS' security and privacy, as well as general end users’ reactions to online scams, research on how IntlS are uniquely impacted by scams remains largely absent.
%

To address this gap, we conduct a two-phase user study comprising surveys (n=$48$) and semi-structured interviews (n=$9$). 
We investigate IntlS' exposure and interactions with scams, post-exposure actions such as reporting, and their perceptions of the usefulness of existing prevention resources and the barriers to following prevention advice.
We find that IntlS are often targeted by scams (\eg attackers impersonating government officials) and fear legal implications or deportation, which directly impacts their interactions with scams (\eg they may prolong engagement with a scammer due to a sense of urgency).
Interestingly, we also find that IntlS may lack awareness of -- or access to -- reliable resources that inform them about scams or guide them in reporting incidents to authorities.
In fact, they may also face unique barriers in enacting scam prevention advice, such as avoiding reporting financial losses, since IntlS are required to demonstrate financial ability to stay in the US.
%
The findings produced by our study help synthesize guidelines for stakeholders to better aid IntlS in reacting to scams.

\end{abstract}


%
\IEEEpeerreviewmaketitle

\section{Introduction}


Each year, thousands of international students arrive in the United States to pursue education at an American institution.
With over a million international students currently studying in the US~\cite{international_students_population}, their journey involves a systematic procedure. This includes applying for a student visa months in advance, attending an interview at an embassy, and awaiting visa approval before traveling to the US~\cite{f1_visa_application}.

As international students adjust to residence in a new country, they confront substantial challenges beyond language or cultural assimilation and financial obligations. Despite possessing certain rights within the US, their visa status inherently includes various restrictions. For example, they are not allowed to work while studying~\cite{f1_employment}, unless their employment is on campus or through participation in programs such as curricular practical training (CPT). Furthermore, those who do not secure employment within a few months after graduation must leave the US~\cite{opt_rules}. Adherence to these restrictions, among others, is critical, as their violation can lead to termination of their legal status. Recent events underscore the impermanence of student visas; specifically, in 2025, international students have experienced numerous instances of unexplained visa revocation. This compels them to undertake serious measures, \eg self-deportation, or in some cases, seeking legal representation~\cite{visa_revocation,chinese_Student_visa_revocation}.

The stress and anxiety that stem from current events and the imperative to meticulously adhere to visa regulations have positioned international students as an at-risk community~\cite{pbs_anxiety,bbc_anxiety}. Prior research has shown that at-risk communities are frequently targeted by threat actors due to their susceptibility to various sociotechnical attacks, including phishing and online harassment~\cite{bellinisok,warfordsok}. This makes international students vulnerable to online scams, where threat actors deceive them into divulging personal information and financial details. Recent events highlight the prevalence of online scams specifically targeting these students, from housing or rental scams~\cite{ctv_rental_scam,dailycardinal_scam} to impersonation of government agencies~\cite{wabe_fbi_scam,et_fbi_warning}. Such incidents can cause victims to disclose sensitive personal details, such as SSNs, and incur financial losses of up to tens of thousands of dollars.

A substantial body of work highlights the devastating consequences of online scams~\cite{button-2014, Houtti_Roy_Gangula_Walker_2024, acharya2024explorativestudypigbutchering, razaq-2021}, however, research that addresses how international students are uniquely impacted by such scams remains absent. Motivated by this, we address this research gap by uncovering the unique scam scenarios they encounter, their interactions with scammers, and the post-scam exposure actions they take. Here, we also gauge the reliability of existing resources available to assist international students in dealing with scams. Lastly, we investigate the barriers that impede them from enacting scam prevention advice or methods. To achieve this, we employ a mixed-methods approach. First, we survey the opinions of 48 international students. We then conduct follow-up semi-structured interviews with nine international students. For both survey free responses and interview transcripts, we qualitatively analyze data, taking an inductive and deductive approach to produce an intermediate codebook, which we iteratively refine. We ceased interviews upon reaching thematic saturation, indicating that we had not overlooked any important themes.

Our methods reveal interesting findings on international students' (IntlS') unique experiences with online scams. We find that IntlS may face difficulty in differentiating scams from legitimate communication, largely due to unfamiliarity with how scams are orchestrated in a different country. They may also inadvertently expose themselves toward targeted scams in order to handle tasks important for their sustenance, such as finding employment or securing housing. Many scammers exploit factors critical to maintaining legal status in the US for IntlS, such as financial self-sustainability, the need to find work soon after the end of the educational program, or insinuations of illegal status or civil violations that could be grounds for, or potentially lead to visa revocation. As a result, IntlS are more susceptible to falling victim to scams, since they may act urgently to handle situations that concern their visa or one of their visa's requirements, and they may not be able to identify scams as easily as someone more familiar with scams common in the US.

However, despite this increased risk and greater potential impact, IntlS find themselves without necessary resources or assistance for dealing with scams. They tend to connect with friends and family as the most trusted source of advice regarding scams. Outside of their personal circle, the closest contact they have are offices in their university, especially the international student office or IT departments. We find that there do not appear to be many tools that can assist them with detecting scams, and many do not know whom or what to consult if they do fall victim to a scam. While some may feel comfortable reporting some scams they encounter, IntlS may be averse to directly seeking out the help of law enforcement, both due to fear of any and all implications that could put their visa status in jeopardy, and uncertainty as to whether they would be able to provide any help at all.

We present our findings, and make recommendations for university bodies to better support their IntlS population, as a more centralized and trusted body to IntlS.
In this paper, we make the following contributions:
\begin{itemize}
    \item  We conduct a mixed-method study to understand how international students uniquely interact with online scams. We survey $48$ international students and conduct $9$ semi-structured interviews. 
    \item We highlight the different scam scenarios international students face, how their experiences as an at-risk user impact their interaction and post scam exposure. We also overview their perceived reliability of scam prevention resources and barriers in enacting prevention advice/resources.
    \item We synthesize key takeaways and recommendations for stakeholders to better prepare international students in dealing with online scams.
\end{itemize}

We make our study's supplementary materials available\footnote{\url{https://osf.io/fx72n/}}.

\section{Background and Related Work}

\subsection{Online Scams and Fraud}
 With the proliferation of digital technology in day-to-day life, users are frequently targeted with innovative nefarious scams, including phishing emails, scam phone calls, ad frauds in mobile apps, cryptocurrency scams, and many others. Researchers have explored the domain of online scams from different perspectives, such as large-scale characterization of online scams, scam detection and prevention methods, and studying experiences of scammed victims to mention a few. In this section, we will discuss prior research on online scams and frauds to situate our work in this space. 
 
\shortsectionBf{Characterizing Scams and Fraud.} To understand the landscape of online scams researchers have resorted to large-scale studies conducted on data collected from publicly available resources, Li et al.~\cite{li-2023} studied the cryptocurrency scam disseminated on social networks popularly known as the `arbitrage bot' scam. This study collected large-scale scam videos from YouTube to analyze dissemination strategies, scale and impact of this scam. Another study~\cite{massimo-2021} systemizes knowledge on online cryptocurrency scams through analyzing data collected from prior literature, public databases and websites. Another large-scale study of online scams~\cite{acharya2024explorativestudypigbutchering} analyzes data from social media, news outlets, and public reports to characterize victims of pig-butchering scams. Although this group of works is important to understand the broad characteristics of online scams, they miss out on personal experiences of users and a qualitative perspective.  

\shortsectionBf{Detection and Prevention Methods.} 
Another direction of research in the online scamming space explores automated detection of scams, and suggests prevention mechanisms. 
Miramirkhani et al.~\cite{Miramirkhani-2017} deployed a crawling infrastructure to detect and collect information on scammers posing as tech-support to users. As part of this work they also interacted with 60 scammers in order to recommend countermeasures. Dong et al.~\cite{dong-2018} designed FraudDroid that  detects ad fraud in Android apps through dynamic analysis of apps. The Surveylance~\cite{kharraz-2018} framework automatically detects online survey-based scams using machine learning methods.       

Researchers have also proposed solutions to assist users in determining whether certain communication is a scam. Franz et al. outlines several categories of these types of intervention methods~\cite{franz2021}. Aside from educating and training users \cite{redmiles-2016, wash-2018}, which are more human-oriented approaches, methods include various UI design decisions to ``nudge'' the user to indicate suspicion or danger of a scam or generating user-comprehensible safety reports\cite{zheng-2023, althobaiti-2021}. Some works also investigate the usefulness of scam-prevention knowledge and resources available for users \cite{redmiles-2020,mossano-2020} and find them to be somewhat lacking in assisting users. These studies show that available guidelines for scam prevention can be abstract, contradictory or difficult to prioritize for users. 

\shortsectionBf{User Experiences with Online Scams and Fraud.}
In addition to the large-scale studies discussed before, researchers have also conducted user surveys and interviews to answer specific research questions centered around online scam experiences. Button et al. \cite{button-2014} conducted interviews and focus groups with online fraud victims and professional stakeholders to understand the reasons people fall for online scams. A recent survey of 8369 participants \cite{Houtti_Roy_Gangula_Walker_2024} from 12 countries discovered global scam patterns across nations -- users from less affluent countries are more susceptible to financial loss and Gross National Income (GNI) per capita is strongly associated with specific scam types. 
More focused works also exist. 
For example, a focused survey on mobile-based scam ecosystem stakeholders in Pakistan \cite{razaq-2021} helps us understand prevalent scam tactics, victims' mental model and possible intervention mechanisms, whereas  Bidgoli et al.~\cite{bidgoli2016cybercrimes} conducts focused interviews to investigate how online scams (among other cybercrimes) impact undergraduates. 

In contrast to these prior works, we focus on characterizing the experiences and challenges of a specific user group -- international students -- with online scams. 
We complement prior works by showcasing how international students face unique challenges that can impact their experiences with scams and how they respond to these scams.

\subsection{At-risk Users}

\mr{
 As defined in prior literature, \cite{bellinisok,warfordsok,matthews2025supporting} ``at-risk users'' are those who face greater chances of being targeted or suffering harm from any kind of attack on their digital safety. This greater risk is often caused by a myriad of external factors, ranging from societal factors (\eg marginalization, social norms) to relationship and personal factors (\eg the user is reliant on the attacker or some other third party, or they require access to a sensitive resource)~\cite{warfordsok}. 

Prior work has shown that different marginalized populations face unique security and privacy risks depending on their social positioning and access to support systems.
For instance, sex workers avoid contacting law enforcement due to fear of criminalization~\cite{mcdonald2021s}, while refugees also face unique security and privacy challenges~\cite{simko2018computer}, dealing with attacks differently as a result~\cite{arunasalam2024understanding}. 
Undocumented immigrants adopt ``low profile strategies'' to avoid institutional scrutiny~\cite{guberek2018keeping}. 
Other groups such as LGBTQ+ individuals~\cite{geeng2022like} and marginalized populations in different regions, \eg Lebanon~\cite{mcclearn2023othered} also face significant security and privacy challenges. 

Another important strand of research highlights how cultural context (especially in non-WEIRD contexts) shapes individuals security and privacy. 
For example, shared-device use in regions such as Bangladesh~\cite{ahmed2017digital} underscore how Western notions of individual device ownership fail to capture practices in regions like the Global South. 
Work with children and families similarly shows how context matters: children often misunderstand online privacy risks~\cite{zhao2019make}, while parents and teens negotiate oversight together~\cite{akter2022parental}. 
These studies demonstrate that risk is not universal.

Within at-risk user research, another line of work emphasizes structural and resource constraints. For example, political campaign workers recognize their roles make them prime targets, but de-prioritize security due to financial and time constraints~\cite{consolvo2021wouldn}.
Similarly, victims of human trafficking face extreme conditions and must adopt specialized safety protocols~\cite{stephenson2025digital}. 
Survivors of intimate partner violence encounter similar constraints, as abusers often maintain access to devices~\cite{havron2019clinical}. 
Demographics can also be a contributing factor. Older adults~\cite{deng2025auntie} and those from lower socioeconomic backgrounds~\cite{vitak2018knew} also experience scams that exploit financial precarity/limited familiarity with digital practices.
}

\shortsectionBf{International Students} (IntlS) in particular are a potential group of at-risk users that has been less explored in literature. While not considered full ``immigrants,'' as they only relocate to the United States temporarily and for the sole purpose of attending school~\cite{intlsdef}, the experience they have is very similar to that of a regular immigrant. They must secure transportation to and housing in the country prior to leaving, they must be proficient in English at an adult level, and they need to adjust to living in an entirely new culture. As students in universities, they also have specific social environments and organizations that they can affiliate with, like the school's international student office or clubs.

In the context of security and privacy, Tran et. al~\cite{tranetal} conducted an interview study on the security and privacy concerns of immigrants (ranging from professional workers to international students) in the United States and briefly touches on immigrants exposure to online scams. 
Similarly, Bidgoli et al.~\cite{bidgoli2017} conduct qualitative case studies of two specific scams targeting international students (through analysis of police reports and interviews), and focus on incident reporting on campus. 
In contrast, our work solely focuses on international students (who are different from work-related immigrants and other immigrant groups), while also extending to other facets of scam interaction not particularly explored by prior works -- including international students' interactions with scams (\eg difficulty distinguishing scams, forced scam exposure), post scam actions (beyond just reporting, \eg spreading awareness among friends), and barriers for enacting advice they learn about scam prevention.

\section{Motivation}

IntlS students face unique challenges and experiences when preparing for and moving to the US for school. 
They must secure housing (possibly off-campus), obtain the appropriate visa, secure transportation, possibly learn a new language, and adjust to an entirely new culture. 
These unique experiences create vulnerabilities for scammers to target and exploit (\eg a rental listing scam that IntlS are more susceptible to because they cannot view the property in person to verify the listing's details).
Additionally, recent events in the United States involving visa revocations for international students~\cite{visa_revocation,chinese_Student_visa_revocation} can compound IntlS' vulnerability and impact the way they handle online scams.

We aim to investigate how IntlS interact with scams/scammers, to determine if and how they are uniquely targeted, and how the community and campus organizations around them can better assist them with scam prevention and recovery. 
We aimed to answer the following research questions:
\begin{itemize}
    \item \textbf{RQ1:} What are the different scam scenarios IntlS face?
    \item \textbf{RQ2:} How do IntlS interact with scammers during these scams?
    \item \textbf{RQ3:} What actions do victims take after scam exposure?
    \item \textbf{RQ4:} How helpful or reliable are scam prevention resources available to IntlS?
    \item \textbf{RQ5:} What barriers do exist in enacting advice/methods for scam prevention?
\end{itemize}
\section{Methodology}

We conduct this study in two phases. 
First, we invite potential participants to complete a pre-screener, where eligible participants are invited to complete an online survey. 
Second, participants who complete the online survey are invited to provide their contact details if they wish to be invited for a follow-up semi-structured interview. 
We opted for this two-phase methodology for the following reasons. 
First, the discussion of online scams -- particularly falling victim to an online scam --  is a sensitive topic. 
Hence, we opted for a survey which allows for participants to provide responses if they may not be comfortable verbally interacting with a researcher (\eg over a platform such as Zoom).
However, given our research questions go in-depth into IntlS interactions with scams, semi-structured interviews were necessary. 
Thus, we make semi-structured interviews voluntary and a follow-up component to our initial survey. 

The findings from the survey primarily address \textbf{RQ1} and \textbf{RQ4}, with some preliminary investigation of \textbf{RQ2}, \textbf{RQ3}, and \textbf{RQ5}. 
The follow-up interview allowed for more detailed responses from participants, more thoroughly addressing all the questions, in particular \textbf{RQ2}, \textbf{RQ3}, and \textbf{RQ5}.

\subsection{Participant Recruitment}
\shortsectionBf{Surveys.} 
We recruited survey participants via two methods: (1)~online survey recruitment tool, Prolific~\cite{prolific} and (2)~direct outreach via university campus. 
For campus outreach, we distributed flyers on the researchers' campuses, advertising the survey in classrooms and emailing relevant student organizations. 
%
%
%
\mr{
To determine participants who were eligible for our study, we prescreened participants. 
On Prolific, we leverage Prolific's feature for prescreening. Participants who passed prescreeners were directed to the full survey. 
For campus outreach, participants completed a separate survey form (hosted on Qualtrics).
We distributed the screener by posting flyers and announcements on university and department social and email channels.  

Participants who passed the screener were provided the full survey. 
Survey participants had to be over 18 and on a U.S. student visa (full prescreening questions are in our supplementary materials).
After pre-screening 66 potential participants on Prolific and 160 potential participants via campus recruitment, we recruited 10 participants and 38 participants from these sources respectively (we had a separate three participants whose data was excluded for data concerns as they did not follow instructions, \eg not answering open-text questions with the right content). 
We note that we did not conduct snowball sampling.
}
Participants who completed the survey were each compensated \$5. 
We ensured survey quality by including two attention checks, which participants passed. 
The median time to complete the survey was 12.9 minutes (Q1-$8.6$ minutes, Q3 - $26.3$ minutes)
We note that compensation amount (prorated) was above the average national minimum wage.

Table~\ref{tab:demographics} presents the demographics of survey participants who comprised diverse backgrounds.
A majority of participants had resided in the U.S. for over five years, held at least a bachelor's degree, and identified as being in a computing-related field. 
Participants also primarily had moved from Asia. 
Our sample was evenly split between male and female participants (with one identifying as non-binary and nine preferring not to answer. 
Most participants were aged between 18 and 34 ($21$ participants).

\begin{table}[t!]
  \caption{Demographics of survey participants.}
  \label{tab:demographics}
  \centering
  \setlength{\tabcolsep}{0.5em}
  \def\arraystretch{0.75}
  \resizebox{\columnwidth}{!}{
  \begin{tabular}{lcc}
    \toprule
    \textbf{Demographic Data} & \textbf{N} & \textbf{\%} \\
    \midrule

    \textit{Years Resided in the U.S. ({\small indicates participants arrival to the U.S.})} & & \\
Less than 1 year & $10$ & $20.83\%$ \\
    2--4 years &  $12$& $25.00\%$ \\
    5+ years & $17$ &$35.42\%$ \\
    Did not specify & $9$ & $18.75\%$ \\

    \midrule
    \textit{Education Level (Completed)} & & \\

    High school diploma &$8$& $16.67$\\
    Some college & $4$ & $8.33\%$ \\
    Bachelor's degree &$14$ & $29.17\%$ \\
    Post-graduate degree (Masters/PhD) & $13$ & $27.08\%$ \\
    Did not specify &$9$ & $18.75\%$\\

    \midrule
    \textit{Region Before U.S.} & & \\
    North America & $3$ & $6.25\%$ \\
    South America &$3$ & $6.25\%$ \\
    Europe & $1$ & $2.08\%$ \\
    Middle East & $3$ & $6.25\%$ \\
    Asia & $21$& $43.75\%$ \\
    Africa &$4$& $8.33\%$ \\
    Australia/Oceania &$2$ & $4.17\%$ \\
    Did not specify & $11$ &$22.92\%$ \\

    \midrule
    \textit{Field in Computing or Related Area} & & \\
    Yes & $27$ & $56.25\%$ \\
    No &$13$ & $27.08\%$\\
    Did not specify & $8$ & $16.67\%$ \\

    \midrule
    \textit{Gender} & & \\
    Male & $19$ & $39.58\%$ \\
    Female &$19$ & $39.58\%$ \\
    Non-binary & $1$& $2.08\%$\\
    Prefer not to answer &$9$ & $18.75\%$ \\

    \midrule
    \textit{Age Group} & & \\
    18--24 & $21$& $43.75\%$ \\
    25--34 & $17$& $35.42\%$\\
    35--44 & $1$& $2.08\%$ \\
    Prefer not to answer &$9$ & $18.75\%$ \\

    \bottomrule
  \end{tabular}
  }
\end{table}

\shortsectionBf{Semi-Structured Interviews.} 
We recruited participants who had completed our online survey for semi-structured interviews. 
\mr{
At the end of our survey, we included a question that prompted participants to check a box if they were willing to be invited for a follow-up interview. 
From the participants who selected that they were willing to be invited, we considered participants a candidate if they indicated having in-depth interactions with a scam that they encountered (\eg conversing with the scammer, responding to a scam email, clicking on a scam link). 
We had seven candidates for interviews from Prolific. We attempted to reach out via Prolific's internal messaging system, but any attempts to reach out to them went without response. 
From campus outreach, we had $17$ candidates. We reached out to them via email and were able to schedule nine interviews. 
Similar to surveys, we did not conduct snowball sampling.
}
We ceased recruitment after reaching thematic saturation of data (detailed in Section~\ref{subsec:data_analysis}).
Participants who completed our interview were paid a $\$10$ Amazon gift card. 
Each interview took around $45$-$60$ minutes, was conducted by 2 authors over Zoom, audio recorded (after consent), transcribed, then anonymized for data analysis. 

\begin{table}[t!]
  \caption{Interview participants demographics.}
  \label{tab:interview_demographics}
  \centering
  \setlength{\tabcolsep}{2em}
  \def\arraystretch{0.75}
  \resizebox{\columnwidth}{!}{
  \begin{tabular}{lcc}
    \toprule
    \textbf{Demographic Data} & \textbf{N} & \textbf{\%} \\
    \midrule

    \textit{Years Resided in the U.S.} & & \\
    Less than 1 year & 2 & 22.22\% \\
    2--4 years & 7 & 77.78\% \\

    \midrule
    \textit{Current Education Level} & & \\
    Undergraduate & 2 & 22.22\% \\
    Master's & 4 & 44.44\% \\
    PhD & 3 & 33.33\% \\

    \midrule
    \textit{Region Before U.S.} & & \\
    Asia & 9 & 100.0\% \\

    \midrule
    \textit{Gender} & & \\
    Female & 7 & 77.78\% \\
    Male & 2 & 22.22\% \\

    \midrule
    \textit{Age Group} & & \\
    18--24 & 3 & 33.33\% \\
    25--34 & 6 & 66.67\% \\

    \bottomrule
  \end{tabular}
  }
\end{table}

Table~\ref{tab:interview_demographics} presents the demographics of interview participants. 
All nine participants came from Asian countries. 
The group was predominantly female, with most participants aged $25$–$34$ and pursuing graduate-level education.
Most participants had resided in the U.S. for less than four years.

\subsection{Ethical Considerations}

Our study and recruitment procedure were reviewed and approved by relevant IRBs (two institutions were involved and each institution's IRB independently approved the study). 
Because our study concerns personal experiences with scams, including potential experiences where participants may have fallen for scams, we allowed participants to skip questions that they may have found embarrassing or sensitive. 
All survey and interview data was stored in a secure storage that only the authors had access to, and any personal information such as the participants' names and universities were anonymized. 
Participant emails were only used for survey and compensation distribution, and were only shared between the authors and their grant manager.

\subsection{Survey and Interview Procedures}

\shortsectionBf{Surveys.}
Our survey was divided into four sections: (1)~experiences with scams, (2)~experiences with anti-scam help and resources, (3)~attitudes towards scams and (4)~demographics. 

The first section focused on the participants' personal experiences with scams. 
We asked them to recount the details of a particularly memorable scam encounter: what happened, what platforms the scammer used, and how they felt and decided to act. 
We also asked them about their general knowledge of scams, and what resources are available to them to give them advice or assistance regarding scams. 

The second section focused more on advice, tools, and resources that participants knew of to assist in detecting, avoiding, and recovering from scams. 
We asked if there were particular organizations or people that the participant knew about that could help them in this manner, and how helpful they have been to them. 
Also in this section, we showed the participants examples of different technological solutions to scam detection or establishing a trustworthy web source, and asked them to indicate how helpful these measures would be to help them avoid scams.
There were 12 examples, gathered from various sources in literature and grouped into categories based on a condensed version of Franz et. al's~\cite{franz2021} taxonomy of user-oriented phishing interventions (barring the training category, as we only consider technological solutions for this particular section): active warnings, passive warnings, visual elements, and color coding. 
There were three examples for each category.
Each participant saw four out of 12 examples (chosen at random, one for each intervention type). 

The third section asked participants about their general attitude toward scams, and how confident and knowledgeable they feel in avoiding them. 
We presented them with several statements about how concerned they are with being scammed, and how they felt current resources and knowledge on scams were helping them. 
We asked them to rate their agreement with the statements on a $5$-point Likert scale (Strongly agree to Strongly disagree). 
We conclude with asking participants demographic questions and allowing participants to check a box to indicate if they were willing to participate in a follow-up interview. 

\shortsectionBf{Interviews.}
We begin interviews by asking participants demographic questions. 
We note that we re-asked demographics as a warm-up before the subsequent part of the interview (they did not have an impact on study outcome).
We requested for consent to record the interview for transcription -- interviews were only recorded if participants provided consent. 
Afterwards, we proceeded with the main questions in the interview. 
Participants were first asked to describe in detail their experience with a particularly memorable scam, their mentality and emotions while dealing with the scam, what they decided to do afterwards, and what may have prevented them from enacting any anti-scam advice in the moment.
Questions in the interview expanded upon the questions asked in the online survey, allowing us to probe for more details about the scammers and the participant's actions. 
We asked about how they felt in the moment encountering the scam, and what motivated them to take the course of action they did. 
We also asked about what they decided to do in the aftermath, if they chose to report the scam to authorities or to alert their personal contacts.
We also took care to avoid leading questions in follow-up questions.

\subsection{Data Analysis}
\label{subsec:data_analysis}
To analyze data, we primarily relied on qualitative analysis~\cite{braun_clarke_2022}. 
We leveraged our defined research questions to ground deductive coding - where we focused on scam scenarios, interactions with scammers, post scam exposure actions, reliability of existing scam prevention resources, and barriers in enacting advice.
We also inductively analyzed data to extract any additional themes.

\shortsectionBf{Surveys.}
Free responses were deductively and inductively coded by two authors who met to reconcile differences. 
Authors produce an intermediate codebook, which is refined after several iterations through this process.
For the survey questions on the helpfulness of technological solutions for scam mitigation, we use a Fisher's Exact Test (with Monte Carlo estimation) to verify if there is any association between intervention type and user reported perceived helpfulness. 
\shortsectionBf{Interviews.}
Interview transcripts were transcribed automatically using Zoom's in-built transcription feature. 
Transcripts were deductively and inductively coded by two authors who familiarized themselves with transcripts and also met to reconcile differences. 
We produced an intermediate codebook that we refined iteratively with each new coding of a transcript.
Here, we note that qualitative analysis was an ongoing process conducted after each interview. We ceased recruitment after reaching thematic saturation~\cite{saunders2018saturation}, indicating that no new themes had emerged (at nine interviews). 
We merge themes extracted during qualitative analysis of our survey free responses with themes from our interviews.

\section{Findings}
In this section, we present our findings from the online survey and interviews. 
We first overview the kinds of scams the participants described.
Afterwards, we detail how the participants interacted with and reacted to these scams. 
Then, we discuss their actions taken after encountering the scam, and detail results pertaining to tools and resources that the participants knew of to help them avoid, detect, and recover from scams. 
Finally, we discuss barriers that the participants experienced that made enacting on prior scam knowledge and seeking help from others difficult.

\subsection{Scam Scenarios}

\begin{table}[t!]
  \caption{Scam platform and scam subject.}
  \label{tab:scam_attributes}
  \centering
  \setlength{\tabcolsep}{2em}
  \def\arraystretch{0.75}
  \resizebox{\columnwidth}{!}{
  \begin{tabular}{lc}
    \toprule
    \textbf{Scam Attributes} & \textbf{N (\%)} \\
    \midrule

    \rowcolor[gray]{0.9} \textit{Scam Platforms} & \\
    Email & 24 (50.0\%) \\
    Text message (SMS) & 24 (50.0\%)  \\
    Phone call & 23 (47.9\%) \\
    Social media & 11 (22.9\%)\\
    Instant messenger app & 3 (6.3\%) \\
    Marketplaces, ``classified sections'', etc. & 3 (6.3\%) \\
    Other websites & 1 (2.1\%) \\

    \midrule
    \rowcolor[gray]{0.9} \textit{Scam Subject} & \\
    Work opportunity & 21 (43.8\%)  \\
    Finances & 20 (41.7\%) \\
    False alerts about gov’t forms & 15 (31.3\%) \\
    Assistance with gov’t forms & 12 (25.0\%) \\
    Romance & 6 (12.5\%) \\
    Social security & 5 (10.4\%) \\
    Rental posting & 3 (6.3\%) \\
    Other & 10 (20.8\%) \\

    \bottomrule
  \end{tabular}
  }
\end{table}

\subsubsection{Overview}
In both the survey and interviews, participants discussed a wide variety of scams.
Table~\ref{tab:scam_attributes} overviews survey participants' responses for scam platforms and subjects  (chosen from a predefined list, with an option to specify free-text responses).  
A majority of the scams received are on very common platforms: text messages, emails, and phone calls. As these are important modes of communication, this finding suggests   that scams are unavoidable. For social media, the most commonly brought up platform was Facebook, either the regular social media platform, groups, or its marketplace (22.2\% of all reported scams, 90.9\% of reported social media scams). 
Interestingly, $25$ participants experienced multi-channel scams (the scam took place on several platforms).

Participants have encountered scams relating to jobs and financial matters the most (43.8\% and 41.7\% respectively), followed closely by scams concerning governmental forms (\eg offering assistance to expedite visa process, supposed errors on forms, 25.0\% and 31.3\% respectively). 
Many participants expressed familiarity with scams in general and how scam communication is worded and presented. Some brought up multiple encounters with scams, some of which were directed against them, and others that affected their friends and family, suggesting that encounters with scams are commonplace and fairly frequent. 

\subsubsection{The Most Memorable Scams of Participants}
When asked to describe one particularly memorable scam, over half of the scams reported by the participants concerned matters of money, or were attempting to steal money from them (52.1\%). These types of scams came in many different forms. Common formats include the scammer claiming the victim had an unpaid toll and providing a link to enter their payment information, or the scammer claiming that the victim had won a prize, and needed to provide payment information or to pay a fee for ``processing.''
%

%

Another common category of reported scams is supposed work opportunities. Scammers appear to use a variety of different jobs to entice victims. \textbf{P34} mentioned receiving a scam for a job at Shein, a large fashion e-commerce platform, \textbf{P45} mentioned receiving a scam from someone claiming to be an unspecified ``well-known company,'' and \textbf{P48} said that they received a scam message offering a government job.

The last common category of memorable scam involves the victim receiving some kind of false alert regarding government forms and matters. A more minor example would be the aforementioned toll scams, as tolls are paid out to government departments of transit.
\mr{ In fact, such toll-text scams have been documented widely~\cite{toll_scam}, with state DMVs warning users that texts demanding toll payments are fraudulent. 
Although toll-text scams target everyone, we found that IntlS were uniquely affected and have differing reactions, since the messages often appear to come from an authority or government entity, which can heighten fear of penalties/visa implication and pressure quicker compliance. 
}

On the more pressing end would be scams pertaining to critical and sensitive documents, like SSNs, for instance. \textbf{P19} received a scam message sent to their university email address, thinking that it was official information pertaining to their new SSN, which they were expecting. They ``didn't expect it to be a scam, since [they] thought nobody can have access to [their university] email address.'' Their account was compromised as a result of this scam.

\begin{table}[t!]
  \caption{Scam tactics reported by participants.}
  \label{tab:scam_tactics}
  \centering
  \setlength{\tabcolsep}{2em}
  \def\arraystretch{0.75}
  \resizebox{\columnwidth}{!}{
  \begin{tabular}{lc}
    \toprule
    \textbf{Scam Tactic} & \textbf{N (\%)} \\
    \midrule
    Scammer attempts to extract money & 25 (52.1\%) \\
    Scammer applies pressure to act/ threats & 12 (25.0\%) \\
    Scammer offers incentive & 12 (25.0\%) \\
    Scammer pretends to be a government body & 11 (22.9\%)\\
    Scammer creates a sense of urgency & 8 (16.7\%) \\
    Scammer pretends to be a personal contact & 4 (8.3\%) \\

    \bottomrule
  \end{tabular}
  }
\end{table}

\subsubsection{Tactics of Scammers}
Table~\ref{tab:scam_tactics} presents the various tactics leveraged by scammers to compel participants to comply with the scam. As mentioned previously, many of these scams had to do with money. We note that not only is stealing money a common goal for scammers, it is also of significant importance to a victim. This is especially so for IntlS, who are abroad and have to monitor their finances to be able to pay for expenses and maintain a visa requirement of financial self-sufficiency. Therefore, our participants indicated that scams involving money would be more likely to be acted upon.

Another common tactic reported was scammers pretending to hold some kind of authority over the participant, or the scams they send imply negative consequences if the victim does not comply. Often, the scammer poses as an employee of some government agency. Sometimes they can be local, as \textbf{P20} described with a scammer that posed as an official from a city's election office. They did not specify what exactly the scammer was looking for, but they asked the participant for their address, and threatened to ``come to [their] house if [they weren't] going to pick up [their] call.'' Toll scams also use this tactic, using the municipal or state government as their mimicked authority. Other times, the scammer poses as a federal official, calling from departments that are relevant to IntlS. 
\mr{ \textbf{P9} discussed a scam where they received a phone call from someone claiming to be from US Immigrant and Customs Enforcement (ICE), which we note is a scam type specifically targeting non-US citizens/green card holders.
}
\begin{quote}
    \textbf{P9:} \textit{It was this person who told me that they were calling from ICE. And then, they just confirmed my identity. They asked if they were speaking to [P9], they asked if I go to [P9's university]. I don't remember if they asked any other identifying questions, but you know, they had it right. And this is where I don't have complete recollection of what they asked, but they told me that something was missing in some database somewhere that I hadn't informed either the school or ICE about something. I don't remember...but it sound like something official.}
\end{quote}

Authorities that scammers impersonate may also be non-government related as well. \textbf{P5} described a scam where they received a text from a scammer pretending to be their academic advisor, who urged them to buy Apple gift cards for them, pressuring them by saying that ``they were in a very important meeting.'' By utilizing victims' superiors and authorities, scammers position their victims to be more likely to do as they say, exploiting trust and power dynamics. 

On the other hand, we found the other common tactic is for the scammer to offer some kind of incentive, be it a job opportunity or a prize of some sort. Multiple participants described scams of this nature. One instance of a job opportunity scam was mentioned by \textbf{P46}. 
\begin{quote}
    \textbf{P46:} \textit{I was scammed [out] of a job opportunity outside USA. They needed some amount claiming it was a fee. I send [it to] them since they really looked convincing. They shared with me photos of the kind of job I wanted. Little did I know it was a scam.}
\end{quote}
An instance of a prize scam was described by \textbf{P3}: 
\begin{quote}
    \textbf{P3:} \textit{They used a web ad to show up on the Lyft app and claimed to provide coupons for Lyft. I thought it was official so I download it and paid for its membership, but later I found they didn't give me the coupon and charged me with monthly membership fees.}
\end{quote}
Our participants noted that by offering something enticing, like a desirable job or a prize of monetary worth, scammers were able to successfully persuade some IntlS into playing along with the scam, and stealing money from them. 

One particularly standout method was mentioned by \textbf{P26}, akin to the act of ``typosquatting,'' or claiming a domain name that is extremely similar to a well-known and trusted domain, in hopes of catching users that accidentally mistype the URL and interact with a phishing site. However, this scam was done with phone numbers instead. 
\begin{quote}
    \textbf{P26:} \textit{So we had a Target Plus Subscription, and it was under my name, so I called to cancel it. Target gave me a number that I should reach back out to, because it was partnered with ShipIt. I accidentally, instead of the number they gave me [that had an] 8, I put in 7, and it went to a scam center. Literally one digit away. }
\end{quote}
The scammers at the wrong number were posing as Target customer service agents and attempted to get \textbf{P26}'s information. \textbf{P26} was confused and did divulge some personal information: their local Target store, their name, and their email. However, when the scammer asked for their subscription's ID number, they realized that they were speaking to a scammer, since if they were truly from Target, they should already have all the information that they were asking \textbf{P26} for. They hung up at this point.

\subsection{Interaction of Participants with Scammers}
\begin{table}[t!]
  \caption{Participants' actions with past scam encounters.}
  \label{tab:participant_actions}
  \centering
  \setlength{\tabcolsep}{2em}
  \def\arraystretch{0.75}
  \resizebox{\columnwidth}{!}{
  \begin{tabular}{p{0.75\columnwidth}c}
    \toprule
    \textbf{Action Taken} & \textbf{N (\%)}\\
    \midrule
    Deleted a scam email/message without interacting further & 38 (79.2\%)  \\
    Hung up a call upon immediately realizing it was a scam & 37 (77.1\%) \\
    Picked up a call from an unknown caller & 35 (72.9\%) \\
    Opened a scam email/message & 34 (70.8\%) \\
    Talked with a scam caller, but hung up when scam was identified & 32 (66.7\%) \\
    Responded to a scam email/message, but stopped when scam was identified & 21 (43.8\%)  \\
    Clicked on a link on a scam email/message & 16 (33.3\%) \\
    Account was compromised because of a scam & 12 (25.0\%) \\
    Opened attachment on a scam email/message & 11 (22.9\%) \\
    Lost personal information, money, or other assets because of a scam & 9 (18.8\%) \\
    \bottomrule
  \end{tabular}
  }
\end{table}

\subsubsection{Overview} 
Table \ref{tab:participant_actions} shows how participants have reacted to any scams they have received in the past. Almost all have either deleted or ignored scam communication, or had minor interactions with them (\eg reading a scam email, picking up and hearing scam caller) before stopping once they realized that it was a scam. Around half (43.8\% for emails/texts, 72.9\% for phone calls) had further communication with the scammer before stopping. Either they had a substantial conversation with a scam caller, or exchanged at least one more message with the scammer than the initial message. Around a third actively interacted with online scams, either by opening an attachment on a scam message (22.9\%) or by clicking on a link in a scam message (33.3\%). 25.0\% of participants reported their account being compromised as a result of a scam, and 18.8\% reported having lost some personal information, money, or other assets because of a scam.  

\subsubsection{Reactions}
Many participants did not have a distinct reaction when encountering the scams they discussed, particularly with email scams. They were able to recognize signs of a scam like ``[job] recruiters don't send text messages to your personal number'' (\textbf{P13}), or ``[asking] for credit card information over the phone'' (\textbf{P6}), or they recognize entire scam schemes, like ``[pretending] they text the wrong number but continue texting for more communication and eventually go money-related'' (\textbf{P7}). Some were skeptical about the scammer's communication and searched further for more information on the scammer, or to verify with a trusted source. \textbf{P45} received an email from someone claiming to be from a ``well-known company,'' but realized it was a scam when they contacted the company's official customer support. \textbf{P5} received a text from a scammer claiming to be their academic advisor, and they recognized that ``this is highly irregular and not something [their] advisor would ever ask of [them],'' but they still messaged their advisor to verify.

A few expressed reactions of fear upon receiving the scam. After receiving a scam email, \textbf{P21} ``was really scared, especially considering the scammer had so much personal information about [them].'' After searching online and finding that it was a scam that was happening to multiple people, they still took action and changed credentials on their accounts, since they were ``worried that the information would be compromised.'' \textbf{P35} was sent a scam claiming that they had an unpaid toll, and since they did truly have an unpaid toll, they clicked on the link the scammer provided. However, they quickly realized that it was a phishing site, and they ``got scared and deleted everything and changed passwords.''

\subsubsection{Difficulty Distinguishing Scams}
Participants found that they sometimes have difficulty telling apart what is a scam and what is legitimate. In fact, sometimes in order to receive important information, they need to receive communication from unknown senders, potentially exposing themselves to scams. A common example discussed by multiple participants was phone scams. When they were applying for jobs and waiting to hear from recruiters, they had to pay attention to calls from unknown numbers. As one participant stated:
\begin{quote}
    \textbf{P31:} \textit{I did a co-op...and [a phone call was] how the recruiter reach back to say that `Hey, you got the internship!' It was a random number and they were like, `Hey, I'm calling from my work phone...' So they tell you to be on the lookout for random call, so you never know [if] it's either a spam call or a recruiter trying to reach you. [...] My paranoia of accidentally missing a recruiter call is high enough to the point where I will still pick up random calls.}
\end{quote}
Another participant stated that they typically report scams they encounter as junk, but they are still concerned about losing real messages. They leveraged verifying the phone area code as a method of determining the trustworthiness of an incoming call or text.
\begin{quote}
    \textbf{P16:} \textit{So whenever I get any call or text with [my local area] code, I respond to them, like my doctor's appointment or my driving lesson appointment, or anything. Because I'm getting a lot of scam messages, I feel like I will just report junk to every message. But in that case I'm losing the authentic messages. So whenever I get this [local area] code, I try to respond to that.}
\end{quote}
Other times, the scammer's communication, links or attachments look realistic enough to seem legitimate and make them difficult to discern as scams. \textbf{P2} described a scenario where the scammer sent a text claiming that they had an Amazon delivery that had arrived, and that they needed to follow a link to receive a PIN code to retrieve it.
\begin{quote}
    \textbf{P2:} \textit{It seemed so realistic, like it was exactly [how it went] when you receive from Amazon. And since Amazon is like a well-recognized website already, you wouldn't take [a look] twice on it, like `Oh, is it really that?' because you wouldn't expect that from a big entity...}
\end{quote}
Still, \textbf{P2} was skeptical and did not interact with the link and verified that this was a scam when they confirmed with friends who shared their Amazon account that no one had actually placed an order.

Another example is a rental scammer discussed by \textbf{P31}, where the scammer managed to obtain an apartment listing site's ``verified'' status.
\begin{quote}
    \textbf{P31:} \textit{I think it might have been apartments.com, and there was an apartment which...in an expensive area, it was one of the cheaper apartments. And I was like...maybe this is a steal, right? [...] Everything made sense, it checked out, and it even had like, `This is a verified lister.' So I was like, `Oh okay, so this has to be a genuine person.'}
\end{quote}
While \textbf{P31} did not suffer any losses from this scam, it was only after calling the scammer, having a suspicious call, receiving a suspicious lease form, and consulting with their boyfriend and their father, that they realized that this was a scam. Having the supposed measure of verification on the listing significantly impacted their interactions with the scam, and why they didn't initially realize it was a scam.

Other times, the timing of receiving scams, whether intentional or coincidental, can make them seem more realistic, if the scam subject concerns something that the victim was actively looking out for. As mentioned previously, \textbf{P35} received a toll scam when they did legitimately have an unpaid toll. They did not mention whether they felt the timing was intentional or not, but regardless, because of their legitimate toll, they were convinced to act further and clicked on the scam message's link. \textbf{P12} had a similar situation with a US Postal Service scam, since they were expecting an important package containing a new credit card.
\begin{quote}
    \textbf{P12:} \textit{I applied for my [new] credit card, and after a while I got a text, and it said that the USPS cannot send me my package because there is some incomplete payment. And since I was new here, I thought that my credit card won't be delivered if I do not pay, or something like that. So there was a link, and I went to the link and it wanted my payment information where I put in my [old] credit card information...}
\end{quote}

\subsubsection{Current Events}
Throughout early 2025, the US government has revoked thousands of IntlS visas, often suddenly and without explanation \cite{visa_revocation, chinese_Student_visa_revocation}. We found that this was a major source of anxiety amongst IntlS, which impacted their interactions with scams. \textbf{P34} and \textbf{P31} discussed how these events affected their recent experiences with scams. \textbf{P34} notes how the threat of visa revocation pushes IntlS to act quickly whenever any potential issue related to their visa or status comes up, potentially missing signs of a scam.
\begin{quote}
    \textbf{Interviewer:} \textit{Do you think we could potentially factor you being an international student into [being persuaded by scams,] and having to remain really good with your visa status?}
    
    \textbf{P34:} \textit{That is one of the concerns...with all of the news and everything. So if anything comes in, you like `Oh okay, first clear this out. First deal with this.'}
\end{quote}
We found that with the threat of putting their visa status in jeopardy, there comes a unique sense of urgency to act when receiving scam communication, not necessarily from just the scammer, but also the IntlS victim. Acting hastily when dealing with scams puts them at a greater risk of falling victim to them, as they do not pause to consider potential suspicious signs, or to look further into the sender.

\subsection{Post-scam Exposure}
After encountering the scams they discussed, the participants took different courses of action, often depending on what scam they had just dealt with. Some mentioned receiving spam emails, but simply deleted them or left them alone after recognizing that they were spam. Another participant, who also received spam emails and text messages, blocked contacts and deleted the messages, but did not report the scam to anyone else. Others reported spam emails to the email service provider or their university's IT department. 

Those that chose not to report did so for different reasons. \textbf{P31} chose not to report a scam, despite it being directly targeted towards them as an IntlS and causing a fearful reaction, since they feared that given current circumstances, reporting the scam or alerting their peers about it would cause panic and potentially lead to more victims. They reflected that they ``probably should tell people about it now that things have kind of died down, but at the time, it was really scary.'' Another participant (\textbf{P22}) expressed that they left spam messages alone without further action, since they ``really [didn't] want to spend more time dealing with scams.'' However, for many participants, they chose not to report the scam to anyone because the possibility did not occur to them, or they did not know who to report to.

Several participants decided to warn their friends and family about the scam they received. In fact, based on participants' responses, personal contacts appear to be a common source of information regarding scams, especially recent ones that should be watched out for. Information about scams is likely exchanged, so participants inform and are informed about them.

\subsection{Help and Resources}

\begin{table}[t!]
  \caption{Participants' available or known resources for assistance with scams.}
  \label{tab:resources}
  \centering
  \setlength{\tabcolsep}{2em}
  \def\arraystretch{0.75}
  \resizebox{\columnwidth}{!}{
  \begin{tabular}{lc}
    \toprule
    \textbf{Resource} & \textbf{N (\%)} \\
    \midrule
    School's IT organization & 15 (31.3\%) \\
    Friends/family & 7 (14.6\%) \\
    School's cybersecurity organization & 6 (12.5\%) \\
    IntlS office & 4 (8.3\%) \\
    Other employer/school department & 2 (4.2\%) \\
    Local law enforcement & 1 (2.1\%) \\
    Not sure & 1 (2.1\%) \\
    Other & 2 (4.2\%) \\
    No resources & 2 (4.2\%) \\
    \bottomrule
  \end{tabular}
  }
\end{table}

\subsubsection{Technological Solutions}
When presented with the examples of technological scam mitigation in Section 2 of the survey, participants appeared to view all methods as helpful (a Fisher's Exact test with Monte Carlo estimation revealed no significant differences, a small effect size ($0.17$) and $95$ CI-$[0.16,0.29]$).
However, in particular with visual elements (\eg logos indicating trust), perceived helpfulness appears to be more mixed. This could potentially be because out of all the methods shown in this study, visual elements may be the easiest to exploit. The examples we showed the participants are all logos of trusted services or organizations: Verisign, a well-known certificate authority, TRUSTe, a certification process by TrustArc that ensures a website is compliant with legal data privacy standards (\eg HIPAA), and PayPal, a large and widely used online payment platform. \cite{Verisign,TrustArc,Paypal}. However, it is important to note here that these are simply images. A scammer could potentially copy and paste them onto a phishing website to make it seem more legitimate and trusted, thus tricking a user. Unless the victim thinks to ask to see documented proof of certification or verifies that a trusted third-party platform is being used, they may see no issue with the inclusion of the logos.

Participants also did not seem to know of many other technological solutions to help avoid scams. Two participants (\textbf{P2}, \textbf{P34}) mentioned the mobile app Truecaller, which identifies unknown callers and warns against potential incoming scam calls. Several other participants mentioned using basic tools like blocking contacts and reporting spam, but no other external tools were mentioned. 

\subsubsection{People and Organizations}
\label{subsubsec:people}
Table \ref{tab:resources} shows the people and organizations available to them for assistance with scams, according to the survey. We found that personal contacts appear to be the most effective resource. For example, \textbf{P2} and \textbf{P12} discussed how they either heard about or alerted their friends and family about recently encountered scams. In the case of participant \textbf{P12}, contacting their sister about a scam scenario they were currently dealing with allowed them to recognize that they were being scammed.
\begin{quote}
    \textbf{P12:} \textit{[...] My sister's also here in the US, so I actually called her and asked her if I can use her [credit] card so that I can complete the process. That's when she mentioned that it's a scam... If I didn't ask [...] my sister, I would not have understood that it was a scam.}
\end{quote}
Outside of their personal circle, the most frequently mentioned trusted organizations were the participants' university's IntlS or IT departments (Note that though IT organizations were the most known resource, participants' described personal contacts as providing more concrete help/advice). Some noted that these departments would sometimes give scam advice alongside orientation presentations, or send occasional news emails about recently reported scams, alerting them of potential scams and what to do if they encounter them. Several participants (\textbf{P3}, \textbf{P31}) also mentioned reporting certain encountered scams to federal government bodies like the FBI or the FTC, via an online anonymous tip form. However, few mentioned seeking out local law enforcement, whether due to reluctance with dealing with the police, or uncertainty if they were the correct people to consult. For example, \textbf{P3} wondered whether they should have reported one discussed scam encounter to the police, but since the scam occurred partly over XiaoHongShu/RedNote, a Chinese social media site, and the scammer was claiming to be a Chinese IntlS, they weren't sure if their town's police would be able to provide assistance, since it might not be within their jurisdiction. Overall, our interviewed participants indicated that they do not have many external resources available or made known to them. They mostly rely on direct personal contacts, or the university that they are a student of. 

\subsection{Barriers to Enacting Advice and Seeking Help}

We found that IntlS appear to have various reasons and barriers to enacting on their preexisting knowledge of scams, and asking for help regarding them. 

\subsubsection{Lack of Knowledge}

Participants expressed that the process for applying for a student visa is long and complex, and living as an IntlS in the US is in itself a challenge. Having to move to an entirely new country and familiarize oneself with a new language, culture, and sets of procedures requires a lot of knowledge, and lacking that knowledge and experience can be exploited by scammers. For example, because IntlS need to handle multiple sets of important government documents, they may be unfamiliar with what is sensitive information and what should be safeguarded. One participant mentioned safeguarding social security numbers:
\begin{quote}
    \textbf{P2:} \textit{Since you're an international student, it might be really difficult for you to know what is legit and what is not, like you shouldn't give out your SSN once you've gotten it, because these are things we are not familiar with... I feel like the social security number was one that caught me off guard, because come on, honestly, it comes on a piece of paper that you're not allowed to laminate. It seems like a very unserious document to me.}
\end{quote}

\mr{
This extends to scammers' tactics as well.
Some participants noted their familiarity with scams in their home country, but less familiarity with how scams operate in the U.S.
One participant from Bangladesh (\textbf{P12}) recounted a scam where the scammer claimed to be the U.S. Postal Service via text, stating that there was an issue with package delivery and prompting the user to visit a phishing website to enter payment information. 
While USPS text scams have been publicly documented and warned against by the USPS itself~\cite{usps_scam,usps_scam_2}, the participant noted that lack of awareness of the USPS communication methods and norms amplified fear and made the request seem more credible.
}

\begin{quote}
    \textbf{P12:} \textit{[...] it's not like I did not face any scams back in Bangladesh, but that was sort of different, because we do not use the postal service that much there. And when I got a text [in the US] from like a postal office, the postal office is kind of like a government organization... So that's why it didn't click to my mind that it might be a scam... Back in Bangladesh, most of the scams that we face is via email.}
\end{quote}
We found in both cases, it was not a lack of prior knowledge about scams that put IntlS at risk. Regardless of their prior knowledge and experience with scams, IntlS would not be able to apply that in these scenarios without additional knowledge of what sensitive information is and how US government bodies communicate through official channels.

\subsubsection{Fear or Reluctance to Report}
\label{subsubsec:fear_report}
As mentioned in Section~\ref{subsubsec:people}, some participants discussed how they were reluctant to report a scam that they had encountered, for various reasons. \mr{ \textbf{P31} received a phone call scam in early 2025, where the scammer claimed to be from the US Department of Homeland Security auditing IntlS. Although homeland security-related scams have been reported~\cite{homeland}, this instance was a unique type of scam specifically targeting IntlS. 
Here, the attacker was trying to coerce the participant into disclosing their SSN to supposedly prove their legal status. 
It is important to note here that many IntlS were initially unaware of how sensitive the SSN was. 
The experience was alarming to \textbf{P31}, as they recount:
}

\begin{quote}
    \textbf{P31:} \textit{That was terrifying to me... That actually scared me for a while, because I went to bed that night, and I was like, `What if it actually was the Department of Homeland Security?'}
\end{quote}

Yet despite this reaction, they chose not to report this scam, since they ``didn't want to alarm people because of everything that was going on.'' This scam was received during a time where many IntlS were having their visas and SEVIS statuses revoked for expressing certain political views, originating from certain countries, or sometimes for no stated reason at all, an issue that is still ongoing at the time of writing~\cite{Knox}. \textbf{P31} decided that in order not to incite panic, they would not report this scam. However, they expressed in hindsight that it would be good to report a little while after encountering the scam. They also reported finding reports from users on Twitter/X alerting others about scams of a similar nature after the encounter.

Participants seem the most comfortable discussing scams with personal contacts, as shown by various participants mentioning how they discussed details of some scams with their friends and family, and heard reports of scams from them. University organizations are often perceived (and expected to be) to be the most trusted external contact, though only some participants suggested they would seek help from those organizations regarding scams if needed. 

Local law enforcement was not brought up as a potential resource by many participants. In an unscripted follow-up (asked as participants consistently raised visa-related concerns), \textbf{P34} noted that IntlS may even be averse to seeking out the help of local police in the aftermath of scam, whether to report them, or to help recover lost assets if the participant fell victim to one. 
\begin{quote}
    \textbf{P34:} \textit{It's very rare that an international student goes to the police. You're in a new place, you don't have people around you, I don't know the procedure, I don't know what's it gonna take, what's gonna happen. Do you have the time to deal with that thing? And if it's a small amount, it's just like `Oh yeah, it's like \$15. It's just gone.' [...] You're scared as a student. It's a scary world. You don't know what's going to happen.}
\end{quote}
Combined with many discussed scams that pressure the victim into complying because of the implications of criminality or civil violations (\eg proof of legal status, tolls or fines), IntlS would be reluctant to seek out law enforcement for assistance for fear that they would be implicated, which would put them at risk of detainment and/or visa revocation.

\subsubsection{The Impact of Visa Status}

Many parts of an IntlS' experience abroad in the US are heavily dependent on their student visa. In order to maintain their status, they must be a full-time student for every semester they study (i.e. they must take at least a minimum number of courses every semester), they must secure housing for themselves, and must show proof that they are financially self-sustainable for their entire time in the US. Their employment opportunities are also limited; they may not work off campus until after one academic year, and even when they can, they can only take work related to their field of study with limited time-span, and they must apply for permission to work through programs like curricular practical training. After their education ends, they only have a 60-day grace period to either leave the country, or to secure full-time employment and apply for a work visa~\cite{f1_employment,f1_visa_application}. 
Participants note that these factors cause them to be very vigilant and aware of opportunities that will help them attain or maintain legal status, as well as scenarios where they are at risk of losing it. 

Factoring this into their encounters with scams, this creates a unique aspect of vulnerability, even against generic scams that do not target them as IntlS in particular. As mentioned in Section~\ref{subsubsec:fear_report}, any scam that involves something as basic as an unpaid toll can have more of an impact on IntlS, since any criminal or civil violation record could have further negative implications towards their visa status. \textbf{P31} additionally comments on IntlS who regularly go to food pantries, the implications of which can carry over to falling victim to scams as well.
\begin{quote}
    \textbf{P31:} \textit{I don't know how much of this is hearsay, and how of this is actually legal, but essentially, when you want to study in the United States, you prove that you are financially capable of taking care of yourself. [...] So if you were to go to a food pantry, you're essentially telling the government that you are not able to afford food, and if you're not able to afford food in the country, you're not financially responsible anymore. And that technically would be grounds for deportation or visa revocation.}
\end{quote}
Participants noted that if they fell victim to a scam and lost money, reporting on that loss, particularly to law enforcement, could mean that government bodies that oversee IntlS affairs would be made aware. If the loss were severe enough that it put the student's financial stability at risk, it could also put their visa status at risk as well. Not only does this make an IntlS reluctant to seek help in the face of losses, knowing that their legal status can be jeopardized as a result of a simple scam impacts their interactions with them. Participants noted that they may feel more anxious, which could make it easier for them to fall victim when encountering a scam.

Another unique impact is on employment or job opportunity scams. Since employment is such an important aspect of being able to stay in the US after an IntlS' education, they will seek job and internship opportunities, both during their university time and after. As a result, not only are they more at risk for being exposed to scams that lure with promises of lucrative jobs, but they may also be more at risk of falling victim to them. They may be eager to take any offer, without pausing to consider the possibility of it being a scam.

The same applies to rental scams. If an IntlS does not acquire on-campus housing, they must search for off-campus properties, and potentially be exposed to fraudulent postings. An additional factor that increases susceptibility is that IntlS will likely search for housing while they are outside the US, as it is a prerequisite step to prepare for their move. Without being present in the area, they cannot view properties in person to verify details about the posting. \textbf{P22} mentioned that legitimate rental posters were willing to hold video calls to directly show them the property and talk face-to-face. However, using video ``tours'' may still not be a foolproof method. \mr{ \textbf{P3} also discussed encountering a rental scam, where the scammer sent them a pre-recorded video showing off the desired property, complete with narration in Mandarin Chinese, a major language from the participant's home country and one they are fluent in.  We note here that the use of Mandarin by the threat actor suggests that they were targeting IntlS, or at least US-based Mandarin speakers.} \textbf{P3} was unaware that this was a scam until they sent a deposit to the scammer, after which they never heard from them again. The video was potentially stolen from another legitimate rental posting, or the scammer had access to the property, but used it to scam Mandarin-speaking IntlS seeking housing in that area. 

Nearly every major aspect of the process of becoming and being an IntlS to the US has the potential to be exploited by scammers. Even with the most generic scams that even domestic students receive, the unique factor of their continually monitored status as a non-immigrant visa holder influences how participants react to and handle scams.

\section{Discussion}

Through analyzing and discussing the details of scam scenarios received by IntlS, we see a significant unique risk arising from IntlS' particular status and experiences in the US, and lacking support for them in scam prevention, detection, and recovery. 
We now discuss broad themes from our findings, overview our study's limitations and provide recommendations for university officials and offices to better support their IntlS student bodies.

\subsection{Key Findings}

\mr{
\textbf{IntlS are at greater risk of falling victim to scams, since they have trouble differentiating scams from legitimate communication, and they must expose themselves more to scams in order to accomplish important tasks.} 
Our results show that IntlS may have more trouble identifying signs and pointers of scams. 
Scams can look very realistic, and due to cultural and procedural differences between the US and their country of origin, as well as general unfamiliarity with scams in the US, IntlS would have more difficulty differentiating what is real and what is fake without experience. 
In addition, to handle important tasks like finding employment and securing housing, IntlS are forced to communicate with strangers, like job recruiters, landlords and new roommates. 
This puts them more at risk of encountering scams at all, since they cannot simply avoid unknown senders. 
These findings align with broader research showing that cultural background shapes what people treat as sensitive or risky and overall privacy experiences. 
For example, privacy concerns differ significantly between Indian and Saudi Whatsapp users, with some cultures treating credential sharing as common~\cite{dev2020lessons}. 
Similarly, Bangladeshi phone-sharing practices demonstrate how Western models of privacy often fail in Global South context~\cite{ahmed2017digital}. 
Together, these comparisons reinforce the differing cultural norms of IntlS can influence interactions with scams. 

\textbf{IntlS may not know of scam assistance resources that are available to them, and even with existing resources, they may feel averse to approaching them for help.}  
Our results also show that IntlS are mostly reliant on their personal contacts for help with scams, and any others tend to be limited to university bodies, like their IntlS office or IT department. 
While some felt comfortable to send anonymous tips to bodies like the FBI, few considered local law enforcement, and even expressed uncertainty in consulting with them. 
This lack of trusted resources put IntlS more at  risk of falling victim to scams, as they have no one else to consult about any scams they encounter, and recovery from a scam is made much more difficult. 
This echoes prior work on other at-risk groups.
To illustrate, sex workers hesitate to engage law enforcement due to fears of criminalization~\cite{mcdonald2021s}. 
Similarly, refugees avoid seeking help due to toxic online harassment~\cite{arunasalam2024understanding}, while undocumented immigrants often keep a low profile to avoid scrutiny~\cite{guberek2018keeping}. 
In all these cases, distrust of institutions increase reliance on informal/personal networks, amplifying vulnerability.

\textbf{Scams have a unique impact on IntlS because of their visa-dependent status in the US, that must be maintained by a number of critical factors. This increases their susceptibility to scams, the impact of falling victim to scams, and who they choose to reach out to for help.} 
To maintain their legal status in the US, IntlS must make many decisions that account for the restrictions and requirements of their visa. 
Many of the requirements they must fulfill are liable to be targets for scammers, like employment, housing and money. 
While many scams do not specifically target them as IntlS or immigrants, the fact that many scams involve these critical requirements has a unique impact on how IntlS interact with scams. 
They may be compelled to act quickly to handle matters that could impact their visa status and comply with the scammer as if their demands are legitimate. 

In the event they do fall victim to a scam, the losses could potentially have a great impact on their status as well. 
A significant financial loss could endanger their status of financial self-sustainability and turning to law enforcement for help is considered risky (\eg due to current events where many IntlS are experiencing sudden visa revocation and, in some cases, detainment and deportation).
These findings mirror how other populations' structural precarity shapes their security practices. 
For example, intimate partner violence survivors depend on carefully managed digital security interventions~\cite{havron2019clinical}, while marginalized groups in Lebanon must weigh visibility against safety in surveillance-heavy environments~\cite{mcclearn2023othered}. 
Similarly, LGBTQ+ communities, who are often victims of targeted attacks themselves, adopt informal safety networks when mainstream advice feels exclusionary~\cite{geeng2022like}. 
Even political campaign workers de-prioritize security under resource constraints~\cite{consolvo2021wouldn}.
Like IntlS, these groups face external pressures that elevate both the stakes of compromise and the barriers to institutional trust.
}

\subsection{Limitations}
\mr{
Our study has several limitations in terms of the demographics of participants that may limit the generalizability of our findings. 
First, participant recruitment was limited to two large US public universities.
While both are large state-funded institutions, this focus may not capture differences in institutional contexts such as smaller colleges or private universities, which could have had an impact on results. 
For example, smaller universities tend to have less resources. 
Most participants were recruited through campus channels despite wider outreach (through Prolific), which potentially introduces recruitment bias.

Additionally, while our interview sample includes participants from multiple countries of origin (specifically, India, Bangladesh, Indonesia, UAE, China), they are confined to the Asian continent. 
This reflects the fact that students from Asia constitute a large proportion of IntlS in the US, but it does not capture experiences of students from other regions. 
To illustrate, a student from Iran may face unique challenges when interacting with authorities due to the scrutiny their country of origin faces, compared to a student from the UK.

Finally, most of our participants were under the age of $35$, meaning that the perspectives of older IntlS who may return for mid-career education/advanced degrees are underrepresented. 
Their life experiences, financial situations and risk perceptions may differ in ways that could meaningfully impact scam susceptibility and reporting behavior.

However, the concerns raised by participants, such as fears of jeopardizing visa status and challenges with available scam-prevention resources likely extend beyond IntlS captured in our sample. 
We encourage future work such as larger and more demographically representative surveys to validate and extend these findings across a broader range of international student populations. 

}

\subsection{Recommendations}
\mr{
Drawing on our findings, we offer structured recommendations for universities to strengthen scam-prevention and response efforts for IntlS. 

\shortsectionBf{Centralized Scam Resource Repository.} Universities should host a centralized repository of scam-related resources tailored to IntlS.
These resources should consolidate the following. 
First, it should provide clear guidance on safeguarding sensitive information (\eg SSNs). 
Second, it should provide examples of scam messages across platforms (emails, texts, calls).
Third, practical advice on identifying scams, \eg links to anti-phishing training programs are also necessary. 
Because IntlS offices are highly trusted contacts, such a repository has the potential to increase accessibility and consistency of information. 

\shortsectionBf{Real-Time Feedback Channels}. 
Students in our study often turned to family/friends during ongoing scams (\eg consulting a sibling), underscoring the need for immediate, trusted advice. 
Universities could establish real-time support channels such as incorporating scam help services with their IT-help desk.
These would be one potential avenue that would allow IntlS to verify suspicious interactions as they unfold. 
This goes beyond current IT processes, which typically only consider reported scam communication after the interaction between scammer and victim has ended. 

\shortsectionBf{Multimodal Reporting Mechanisms.} In connection to the above, universities should provide multimodal reporting endpoints to capture the full range of scams targeting students and to overcome current limitations (existing IT reporting systems typically only handle email). 
Options could include SMS-forwarding numbers, web upload portals for screenshots, or app-based reporting. 
These systems would broaden institutional awareness of scam tactics and enable relevant bodies such as international student offices to issue timely alerts to the community.

\shortsectionBf{Moderated Peer-Support Spaces.} Complementing our recommendation for real-time feedback channels, we see moderated peer-support spaces as a potential solution.
This builds on the fact that our participants frequently relied on peers for reassurance and advice. Universities can build on this by providing moderated anonymous online spaces where students can share suspicious messages and receive feedback from peers and trained and appointed moderators (who would also be students). 
Such spaces may reduce the isolation many IntlS feel and distribute the emotional burden of scam response more equitably.

\shortsectionBf{Timely and High-Quality Resource Delivery.} We note that the aforementioned recommendations above are useful but alone, they are not sufficient as resource quantity alone is not a clear indication of an appropriate solution. 
Scam tactics evolve rapidly, making timeliness and adaptability critical. 
International student offices should take responsibility for holding periodic seminars on emerging scam trends (as opposed to one-time training), issue regular scam-awareness newsletters/alerts and ensure that staff receive ongoing cyber security training so they can offer accurate, up-to-date guidance. 
This emphasis on resource quality and timeliness ensures support remains relevant rather than static.
Additionally, future work can complement these recommendations in order to develop a better understanding of resource quality. 
For instance, systematic evaluations could assess whether students find newsletters actionable, whether seminars actually improve scam detection confidence, and whether training equips staff to recognize new scam modalities. 
Longitudinal studies could also measure how resource delivery frequency and responsiveness affect student trust and reporting behavior.

\shortsectionBf{Supporting Engagement with Authorities.} 
We found that students expressed fear of jeopardizing visa status when engaging with outside authorities. 
Universities can help bridge this gap by assisting students in contacting law enforcement or other authoritative bodies. 
Having a trusted university representative involved can increase student confidence, reduce risks of miscommunication and ensure scams are reported beyond the campus context.

\shortsectionBf{Considering Unintended Consequences.} While university-based reporting systems are valuable, they may inadvertently discourage students from reporting scams to external authorities.
Thus, universities should frame their systems are complementary rather than substitutive, guiding students toward both internal and external reporting.
}
\section{Conclusion}

With their particular experiences and challenges, IntlS are uniquely at a higher risk of falling victim to scams. The encounters they have with scams can cause significant levels of anxiety, and the losses they suffer can potentially have consequences that extend to their ability to legally remain in the US. Compounded with current events that only add to their risk factor, it is important that IntlS receive the help, knowledge, and training they need to avoid and handle scams without jeopardizing their assets and their safety. Our study addresses the knowledge gap on the particular circumstances of this at-risk group, finding that current resources are not enough to support them. We give recommendations for university bodies like IntlS offices and IT departments to work to provide more support for IntlS. With the findings from this study, we hope to shine a light on this very pertinent issue, and to encourage universities to provide greater support for this critical part of their communities.

\section*{Acknowledgement}
This work is supported by startup funding from University of Massachusetts Amherst and Purdue University.



%
\bibliographystyle{IEEEtran}
\bibliography{references}

\begin{thebibliography}{10}
\providecommand{\url}[1]{#1}
\csname url@samestyle\endcsname
\providecommand{\newblock}{\relax}
\providecommand{\bibinfo}[2]{#2}
\providecommand{\BIBentrySTDinterwordspacing}{\spaceskip=0pt\relax}
\providecommand{\BIBentryALTinterwordstretchfactor}{4}
\providecommand{\BIBentryALTinterwordspacing}{\spaceskip=\fontdimen2\font plus
\BIBentryALTinterwordstretchfactor\fontdimen3\font minus \fontdimen4\font\relax}
\providecommand{\BIBforeignlanguage}[2]{{%
\expandafter\ifx\csname l@#1\endcsname\relax
\typeout{** WARNING: IEEEtran.bst: No hyphenation pattern has been}%
\typeout{** loaded for the language `#1'. Using the pattern for}%
\typeout{** the default language instead.}%
\else
\language=\csname l@#1\endcsname
\fi
#2}}
\providecommand{\BIBdecl}{\relax}
\BIBdecl

\bibitem{international_students_population}
``The {U.S.} has more than 1 million foreign students. here’s who they are.'' \url{https://www.washingtonpost.com/education/2025/05/25/international-students-harvard-trump/}, 2025, [Accessed: 2025-06-01].

\bibitem{f1_visa_application}
``Student visa,'' \url{https://travel.state.gov/content/travel/en/us-visas/study/student-visa.html}, 2025, [Accessed: 2025-06-01].

\bibitem{f1_employment}
``Student and employment,'' \url{https://www.uscis.gov/working-in-the-united-states/students-and-exchange-visitors/students-and-employment}, 2025, [Accessed: 2025-06-01].

\bibitem{opt_rules}
``Optional practical training {(OPT)} for {F-1} students,'' \url{https://www.uscis.gov/working-in-the-united-states/students-and-exchange-visitors/optional-practical-training-opt-for-f-1-students}, 2025, [Accessed: 2025-06-01].

\bibitem{visa_revocation}
``Student visa terminations have quickly hit over half of all states.'' \url{https://www.nbcnews.com/news/asian-america/international-students-revoked-visas-reasons-why-rcna200313}, [Accessed: 2025-06-01].

\bibitem{chinese_Student_visa_revocation}
``Trump administration will ‘aggressively revoke’ {Chinese} student visas in major escalation with beijing,'' \url{https://www.cnn.com/2025/05/28/politics/student-visa-china-revoke-rubio}, [Accessed: 2025-06-01].

\bibitem{pbs_anxiety}
``Visa cancellations and deportations sow panic for international students,'' \url{https://www.pbs.org/newshour/politics/visa-cancellations-and-deportations-sow-panic-for-international-students}, 2025, [Accessed: 2025-06-01].

\bibitem{bbc_anxiety}
``Anxiety at {US} colleges as foreign students are detained and visas revoked,'' \url{https://www.bbc.com/news/articles/c20xq5nd8jeo}, 2025, [Accessed: 2025-06-01].

\bibitem{bellinisok}
R.~Bellini, E.~Tseng, N.~Warford, A.~Daffalla, T.~Matthews \emph{et~al.}, ``{SoK}: Safer digital-safety research involving at-risk users,'' in \emph{IEEE Symposium on Security and Privacy (SP)}, 2024.

\bibitem{warfordsok}
N.~Warford, T.~Matthews, K.~Yang, O.~Akgul, S.~Consolvo \emph{et~al.}, ``{SoK}: A framework for unifying at-risk user research,'' in \emph{IEEE Symposium on Security and Privacy (SP)}, 2022.

\bibitem{ctv_rental_scam}
``International students in {Kitchener, Ont.} lose thousands of dollars to alleged rental scam,'' \url{https://www.ctvnews.ca/kitchener/article/international-students-in-kitchener-ont-lose-thousands-of-dollars-to-alleged-rental-scam/}, 2024, [Accessed: 2025-06-05].

\bibitem{dailycardinal_scam}
``{UWPD} warns students of increased scams targeting international students,'' \url{https://www.dailycardinal.com/article/2024/03/uwpd-warns-students-of-increased-scams-targeting-international-students}, 2024, [Accessed: 2025-06-05].

\bibitem{wabe_fbi_scam}
``{FBI’s Atlanta} office warns of scam targeting international students,'' \url{https://www.wabe.org/fbis-atlanta-office-warns-of-scam-targeting-international-students/}, 2025, [Accessed: 2025-06-05].

\bibitem{et_fbi_warning}
``Us international students warned by {FBI} about scam calls from {US} immigration,'' \url{https://economictimes.indiatimes.com/nri/latest-updates/us-international-students-warned-by-fbi-about-scam-calls-from-us-immigration/articleshow/121205035.cms}, 2025, [Accessed: 2025-06-05].

\bibitem{button-2014}
M.~Button, C.~M. Nicholls, J.~Kerr, and R.~Owen, ``Online frauds: Learning from victims why they fall for these scams,'' \emph{Australian \& New Zealand Journal of Criminology}, 2014.

\bibitem{Houtti_Roy_Gangula_Walker_2024}
M.~Houtti, A.~Roy, V.~N.~R. Gangula, and A.~Walker, ``A survey of scam exposure, victimization, types, vectors, and reporting in 12 countries,'' \emph{Journal of Online Trust and Safety}, 2024.

\bibitem{acharya2024explorativestudypigbutchering}
B.~Acharya and T.~Holz, ``An explorative study of pig butchering scams,'' \emph{arXiv preprint arXiv:2412.15423}, 2024.

\bibitem{razaq-2021}
L.~Razaq, T.~Ahmad, S.~Ibtasam, U.~Ramzan, and S.~Mare, ``"we even borrowed money from our neighbor": Understanding mobile-based frauds through victims' experiences,'' \emph{ACM on human-computer interaction}, 2021.

\bibitem{li-2023}
K.~Li, S.~Guan, and D.~Lee, ``Towards understanding and characterizing the arbitrage bot scam in the wild,'' \emph{ACM SIGMETRICS Performance Evaluation Review}, 2024.

\bibitem{massimo-2021}
M.~Bartoletti, S.~Lande, A.~Loddo, L.~Pompianu, and S.~Serusi, ``Cryptocurrency scams: Analysis and perspectives,'' \emph{IEEE Access}, 2021.

\bibitem{Miramirkhani-2017}
N.~Miramirkhani, O.~Starov, and N.~Nikiforakis, ``Dial one for scam: A large-scale analysis of technical support scams,'' in \emph{Network and Distributed System Security Symposium}, 2017.

\bibitem{dong-2018}
F.~Dong, H.~Wang, L.~Li, Y.~Guo, T.~F. Bissyand\'{e} \emph{et~al.}, ``{FraudDroid}: automated ad fraud detection for {Android} apps,'' in \emph{ACM Joint Meeting on European Software Engineering Conference and Symposium on the Foundations of Software Engineering}, 2018.

\bibitem{kharraz-2018}
A.~Kharraz, W.~Robertson, and E.~Kirda, ``Surveylance: Automatically detecting online survey scams,'' in \emph{IEEE Symposium on Security and Privacy (SP)}, 2018.

\bibitem{franz2021}
A.~Franz, V.~Zimmermann, G.~Albrecht, K.~Hartwig, C.~Reuter \emph{et~al.}, ``{SoK}: Still plenty of phish in the sea {\textemdash} a taxonomy of {User-Oriented} phishing interventions and avenues for future research,'' in \emph{Symposium on Usable Privacy and Security}, 2021.

\bibitem{redmiles-2016}
E.~M. Redmiles, A.~R. Malone, and M.~L. Mazurek, ``{I Think They're Trying to Tell Me Something: Advice Sources and Selection for Digital Security},'' in \emph{IEEE Symposium on Security and Privacy (SP)}, 2016.

\bibitem{wash-2018}
R.~Wash and M.~M. Cooper, ``Who provides phishing training? facts, stories, and people like me,'' in \emph{CHI Conference on Human Factors in Computing Systems}, 2018.

\bibitem{zheng-2023}
S.~Y. Zheng and I.~Becker, ``Checking, nudging or scoring? evaluating e-mail user security tools,'' in \emph{Symposium on Usable Privacy and Security (SOUPS)}, 2023.

\bibitem{althobaiti-2021}
K.~Althobaiti, N.~Meng, and K.~Vaniea, ``I don’t need an expert! making {URL} phishing features human comprehensible,'' in \emph{CHI Conference on Human Factors in Computing Systems}, 2021.

\bibitem{redmiles-2020}
E.~M. Redmiles, N.~Warford, A.~Jayanti, A.~Koneru, S.~Kross \emph{et~al.}, ``A comprehensive quality evaluation of security and privacy advice on the web,'' in \emph{USENIX Security Symposium}, 2020.

\bibitem{mossano-2020}
M.~Mossano, K.~Vaniea, L.~Aldag, R.~Düzgün, P.~Mayer \emph{et~al.}, ``Analysis of publicly available anti-phishing webpages: contradicting information, lack of concrete advice and very narrow attack vector,'' in \emph{IEEE European Symposium on Security and Privacy Workshops}, 2020.

\bibitem{bidgoli2016cybercrimes}
M.~Bidgoli, B.~P. Knijnenburg, and J.~Grossklags, ``When cybercrimes strike undergraduates,'' in \emph{APWG Symposium on Electronic Crime Research (eCrime)}, 2016.

\bibitem{matthews2025supporting}
T.~Matthews, E.~Bursztein, P.~G. Kelley, L.~Kissner, A.~Kramm \emph{et~al.}, ``Supporting the digital safety of at-risk users: Lessons learned from 9+ years of research \& training,'' \emph{ACM Transactions on Computer-Human Interaction}, 2025.

\bibitem{mcdonald2021s}
A.~McDonald, C.~Barwulor, M.~L. Mazurek, F.~Schaub, and E.~M. Redmiles, ``" it's stressful having all these phones": Investigating sex workers' safety goals, risks, and practices online,'' in \emph{USENIX Security Symposium}, 2021.

\bibitem{simko2018computer}
L.~Simko, A.~Lerner, S.~Ibtasam, F.~Roesner, and T.~Kohno, ``Computer security and privacy for refugees in the {United States},'' in \emph{IEEE symposium on security and privacy}, 2018.

\bibitem{arunasalam2024understanding}
A.~Arunasalam, H.~Farrukh, E.~Tekcan, and Z.~B. Celik, ``Understanding the security and privacy implications of online toxic content on refugees,'' in \emph{USENIX Security Symposium}, 2024.

\bibitem{guberek2018keeping}
T.~Guberek, A.~McDonald, S.~Simioni, A.~H. Mhaidli, K.~Toyama \emph{et~al.}, ``Keeping a low profile? technology, risk and privacy among undocumented immigrants,'' in \emph{CHI conference on human factors in computing systems}, 2018.

\bibitem{geeng2022like}
C.~Geeng, M.~Harris, E.~Redmiles, and F.~Roesner, ``"like lesbians walking the perimeter": Experiences of {U.S.} {LGBTQ+} folks with online security, safety, and privacy advice,'' in \emph{USENIX Security Symposium}, 2022.

\bibitem{mcclearn2023othered}
J.~McClearn, R.~B. Jensen, and R.~Talhouk, ``Othered, silenced and scapegoated: Understanding the situated security of marginalised populations in lebanon,'' in \emph{USENIX Security Symposium}, 2023.

\bibitem{ahmed2017digital}
S.~I. Ahmed, M.~R. Haque, J.~Chen, and N.~Dell, ``Digital privacy challenges with shared mobile phone use in {Bangladesh},'' \emph{Proceedings of the ACM on Human-computer Interaction}, 2017.

\bibitem{zhao2019make}
J.~Zhao, G.~Wang, C.~Dally, P.~Slovak, J.~Edbrooke-Childs \emph{et~al.}, ``I make up a silly name' understanding children's perception of privacy risks online,'' in \emph{CHI conference on human factors in computing systems}, 2019.

\bibitem{akter2022parental}
M.~Akter, A.~J. Godfrey, J.~Kropczynski, H.~R. Lipford, and P.~J. Wisniewski, ``From parental control to joint family oversight: Can parents and teens manage mobile online safety and privacy as equals?'' \emph{ACM on Human-Computer Interaction}, 2022.

\bibitem{consolvo2021wouldn}
S.~Consolvo, P.~G. Kelley, T.~Matthews, K.~Thomas, L.~Dunn \emph{et~al.}, ``"why wouldn't someone think of democracy as a target?": Security practices \& challenges of people involved with {U.S.} political campaigns,'' in \emph{USENIX Security Symposium}, 2021.

\bibitem{stephenson2025digital}
S.~Stephenson, L.~Ramjit, T.~Ristenpart, and N.~Dell, ``Digital technologies and human trafficking: Combating coercive control and navigating digital autonomy,'' in \emph{CHI Conference on Human Factors in Computing Systems}, 2025.

\bibitem{havron2019clinical}
S.~Havron, D.~Freed, R.~Chatterjee, D.~McCoy, N.~Dell \emph{et~al.}, ``Clinical computer security for victims of intimate partner violence,'' in \emph{USENIX Security Symposium}, 2019.

\bibitem{deng2025auntie}
Y.~Deng, C.~He, Y.~Zou, and B.~Li, ``"auntie, please don't fall for those smooth talkers": How {Chinese} younger family members safeguard seniors from online fraud,'' in \emph{CHI Conference on Human Factors in Computing Systems}, 2025.

\bibitem{vitak2018knew}
J.~Vitak, Y.~Liao, M.~Subramaniam, and P.~Kumar, ``"i knew it was too good to be true": The challenges economically disadvantaged internet users face in assessing trustworthiness, avoiding scams, and developing self-efficacy online,'' \emph{ACM on human-computer interaction}, 2018.

\bibitem{intlsdef}
``International student life cycle,'' \url{https://studyinthestates.dhs.gov/students/get-started/international-student-life-cycle}, 2025, [Accessed: 2025-07-16].

\bibitem{tranetal}
M.~Tran, C.~W. Munyendo, H.~Sri~Ramulu, R.~G. Rodriguez, L.~Ball~Schnell \emph{et~al.}, ``Security, privacy, and data-sharing trade-offs when moving to the {United States}: Insights from a qualitative study,'' in \emph{IEEE Symposium on Security and Privacy (SP)}, 2024.

\bibitem{bidgoli2017}
M.~Bidgoli and J.~Grossklags, ``“hello. this is the {IRS} calling.”: A case study on scams, extortion, impersonation, and phone spoofing,'' in \emph{APWG Symposium on Electronic Crime Research (eCrime)}, 2017.

\bibitem{prolific}
``Prolific,'' \url{https://www.prolific.co}, 2025, [Accessed: 2025-03-26].

\bibitem{braun_clarke_2022}
V.~Braun and V.~Clarke, ``Conceptual and design thinking for thematic analysis.'' \emph{Qualitative Psychology}, 2022.

\bibitem{saunders2018saturation}
B.~Saunders, J.~Sim, T.~Kingstone, S.~Baker, J.~Waterfield \emph{et~al.}, ``Saturation in qualitative research: Exploring its conceptualization and operationalization,'' \emph{Quality \& Quantity}, 2018.

\bibitem{toll_scam}
``Got a text about unpaid tolls? it’s probably a scam,'' \url{https://consumer.ftc.gov/consumer-alerts/2025/01/got-text-about-unpaid-tolls-its-probably-scam/}, 2025, [Accessed: 2025-10-01].

\bibitem{Verisign}
``Verisign,'' \url{https://www.verisign.com/}, 2025, [Accessed: 2025-07-16].

\bibitem{TrustArc}
``Trustarc data privacy certification standards,'' \url{https://trustarc.com/consumer-information/privacy-certification-standards/}, 2025, [Accessed: 2025-07-16].

\bibitem{Paypal}
``Paypal,'' \url{https://www.paypal.com/us/home}, 2025, [Accessed: 2025-07-16].

\bibitem{usps_scam}
``Smishing: Package tracking text scams,'' \url{https://www.uspis.gov/news/scam-article/smishing-package-tracking-text-scams}, 2025, [Accessed: 2025-10-01].

\bibitem{usps_scam_2}
``Scams \& scheme alerts,'' \url{https://faq.usps.com/s/article/Scams-Scheme-Alerts}, 2025, [Accessed: 2025-10-01].

\bibitem{homeland}
``{Homeland Security} phone scam leaves people scared and confused,'' \url{https://www.idtheftcenter.org/post/homeland-security-phone-scam-leaves-people-scared-and-confused}, 2021, [Accessed: 2025-10-01].

\bibitem{Knox}
``Where international students stand now,'' \url{https://www.insidehighered.com/news/global/international-students-us/2025/06/03/chronicling-trumps-evolving-international-student}, [Accessed: 2025-07-16].

\bibitem{dev2020lessons}
J.~Dev, P.~Moriano, and L.~J. Camp, ``Lessons learnt from comparing {WhatsApp} privacy concerns across {Saudi} and {Indian} populations,'' in \emph{Symposium on Usable Privacy and Security}, 2020.

\end{thebibliography}
\appendices
\newpage 

\appendices 

\section{Meta-Review}
The following meta-review was prepared by the program committee for the 2026
IEEE Symposium on Security and Privacy (S\&P) as part of the review process as
detailed in the call for papers.

\subsection{Summary}
This paper examines the scams targeting US-based international students through a two-phase qualitative study. It highlights several unique challenges international students face that make scam prevention difficult. It also makes concrete defensive recommendations involving trusted university offices.

\subsection{Scientific Contributions}
\begin{itemize}
\item Addresses a Long-Known Issue
\item Provides a Valuable Step Forward in an Established Field
\end{itemize}

\subsection{Reasons for Acceptance}
\begin{enumerate}
\item This paper provides a valuable step forward in an established field by showing (US-based) international students as a specially targeted at-risk population. The paper exposes unique weaknesses of this group such as the need to maintain visa status, lack of cultural knowledge all of which appear to both increase their vulnerability and contribute to under-reporting behavior.
\item The paper also attempts to address this issue by making concrete actionable recommendations involving the active role of international student centers to help in scam mitigation.
\end{enumerate}

\end{document}